# DESCRIPTION OF THE COLLECTIVE STATES IN $^{160}$Dy.

V.P.Garistov[2], A.A Solnyshkin[1], J.Adam[1], V.G.Kalinnikov[1]

[1] Joint Institute for Nuclear Research Dubna, Russia.
[2] Institute for Nuclear Research and Nuclear Energy, Sofia, Bulgaria

## INTRODUCTION

The $^{160}$Dy nucleus is classified as deformed nucleus ($\beta \sim 0.23$) and has quite a complicated scheme of excited states. By now it has been quite well studied experimentally in nuclear reactions, Coulomb excitation, and $\beta$ decays of $^{160}$Tb and $^{160m,g}$Ho [1]. Our recent investigation of the decay $^{160}$Er $\rightarrow$ $^{160m,g}$Ho $\rightarrow$ $^{160}$Dy [2] has made it possible to expand considerably the scheme of excited $^{160}$Dy states and to correlate the reaction and $\beta$-decay data. Over a hundred new levels are added to the previously known excited states in the $^{160m,g}$Ho $\rightarrow$ $^{160}$Dy decay scheme.

In this paper a theoretical analysis of the experimental data of the spectrum of excited states in $^{160}$Dy is presented. The Interacting Vector Boson Model (IVBM) [3] has been applied for classification of low lying collective excited states with $J^{\pi} = 0^+, 2^+, 4^+, 6^+$ and also for description of the ground, beta, gamma and octupole bands energies.

## LOW-LYING COLLECTIVE STATES IN $^{160}$Dy

The spectra of the excited states (164 states) and $\gamma$-transitions (878 transitions) between them in the $^{160}$Dy nucleus are rather complicated (See the decay scheme). It is clear that these experimental investigations are considerably difficult. In our experiments we used two types of sources to study the structure of the excited $^{160}$Dy states populated by the $\beta$-decay. One was an equilibrium mixture of $^{160g}$Er and daughter $^{160m}$Ho and $^{160g}$Ho nuclei and the other was an equilibrium mixture of $^{160m}$Ho and $^{160g}$Ho nuclei. Under these conditions it is impossible to determine intensities of population of $^{160}$Dy states by the $^{160m}$Ho and $^{160g}$Ho decays separately. However, current progress in laser methods for separation of not only isotopes but also isomers seems to allow successful solution of this problem. To this end it is necessary to separate pure $^{160g}$Ho from the mixture of $^{160m}$Ho and $^{160g}$Ho and to carry out spectroscopic measurements.

As a matter of fact these experimental data arouse great interest from theoretical point of view and provoke finding new approaches in description and comprehension of the nuclear structure.

In the recently published papers [4] it has been shown with the use of the simple model Hamiltonian that low-lying collective states of an even-even deformed nucleus can be described with a good accuracy by the parabolic dependence of the energies of these levels on the number of collective excitations

(for example, for each excited $0^+$ state one may put into accordance a number of monopole bosons). In this paper we show that the low-lying excited states in the same nucleus can be also classified by number of collective excitations within the framework of the interacting vector boson model (IVBM) [3]. IVBM is algebraic model with a building block constructed by two types of vector bosons $u_m(\alpha)$ and $u_m(\beta)$ The bilinear combinations of the boson creation and annihilation operators, $u^+_m(\alpha)$ and $u_m(\alpha)$ respectively, generate a non-compact symplectic group $Sp(12,R)$.

Operators

$$F^L_M(\alpha,\beta) = \sum_{k,m} C^{LM}_{1k1m} u^+_k(\alpha) u^+_m(\beta)$$
$$G^L_M(\alpha,\beta) = \sum_{k,m} C^{LM}_{1k1m} u_k(\alpha) u_m(\beta) \qquad (1)$$
$$A^L_M(\alpha,\beta) = \sum_{k,m} C^{LM}_{1k1m} u^+_k(\alpha) u_m(\beta)$$

generate the maximum compact subgroup $U(6)$ of the $Sp(12,R)$ group. The rotational limit of the model is defined by the chain:

$$U(6) \supset SU(3) \otimes U(2) \supset SO \otimes U(1) \qquad (2)$$
$$[N] \qquad (\lambda,\mu) \qquad (N,T) \qquad KL \qquad T_0$$

The labels of the subgroups given beneath the chain are the quantum numbers of the irreducible representations. There is another possibility of decomposition of the $Sp(12,R)$ group in terms of the non-compact $Sp(4,R)$ group in the chain:

$$Sp(12,R) \supset Sp(4,R) \otimes SO(3) \qquad (3)$$

In a new application of the algebraic Interacting Vector Boson Model (IVBM), we make use of the reduction of its group of dynamical symmetry $Sp(12,R)$ through the group $Sp(4,R) \otimes SO(3)$, which defines the basis of states with a fixed value of the angular momentum L. The correspondence of this reduction to the one trough the $U(6) \supset U(3) \otimes U(2)$, giving the rotational limit of the model yelds the possibility to study the energy distribution of the collective states with fixed angular momentum.

The properties of these two $Sp(12,R)$ decomposition chains allow the diagonalization of the Hamiltonian and calculation of the level energies:

$$E[(N,T);KLM;T_0] = aN + bN^2 + \alpha_3 T(T+1) + \beta_3 L(L+1) + \alpha_1 T_0 \qquad (4)$$

For any fixed value of angular momentum L this energy is the simple second order function of number of bosons N:

$$E[N] = aN + bN^2 + c \qquad (5)$$

Using the energy distribution (5) for the L = 0, 2, 4 and 6 we obtain the parameters a, b, and c from the comparison of our calculations with experiment for low lying collective states energies in $^{160}$Dy We present our calculations and comparison with experiment in Figure. The average energy deviations $<| E_{expt} - E_{calc} |>$ are 10.7, 39.4, 21.9, and 50.4 keV for the levels with $J^\pi = 0^+, 2^+, 4^+$, and $6^+$ respectively. With nice accuracy the experimental energies for low lying collective states follow the parabolic distribution function of number of collective excitations.

# DESCRIPTION OF THE GROUND ROTATIONAL BAND, γ-VIBRATIONAL AND OCTUPOLE BANDS IN $^{160}$Dy

In the interacting vector boson model (IVBM) with the symplectic Sp(12,R) group [3] the spectrum of excited states of even-even deformed nuclei is determined by the labels of the subgroup representations in the corresponding decomposition chain of the group.

The expression for the energy in terms of $\{\lambda,\mu\}$ multiplets for the rotational limit chain has the form

$$E[\{\lambda\mu\};KLM;T_0] = \beta_3 L(L+1) + aN + \alpha_1 N(N+5) + \alpha_3(\lambda^2+\mu^2+\lambda\mu+3\lambda+3\mu) + cT_0^2 \quad (6)$$

and for the second chain it is, in terms of $\{N,T\}$,

$$E[(N,T);KLM;T_0] = aN + bT_0^2 + \alpha_3 T(T+1) + \beta_3 L(L+1) + cT_0^2 \quad (7)$$

The main reduction rules are as follows:

N-even → 0, 2, 4, 6...
T = (N/2), (N/2)-1, (N/2)-2,..., 0 or 1;
$T_0$ = -T, -T+1,...T;
λ=2T; μ=N/2-T.
K = min(λ,μ), min(λ,μ) -2,......, 0 or 1.
If K = 0 → L=max(λ,μ), L=max(λ,μ) -2...,0,1.
And if K≠ 0 → L=max(λ,μ), L=max(λ,μ)-1,...,0,1.

The state parity is defined as $\pi = (-1)^T$.

In the (λ,μ) description the ground-state rotational band corresponds to (λ=0, μ=L) and T=0, $T_0$=0, the octupole band is defined by the multiplet (λ=2, μ=L-1), T=1 or (λ=6, μ=L-3), T=3, $T_0$=1, the γ-band will be defined by the multiplets (λ=4, μ=L-2), T=2, $T_0$=1 or (λ=8, μ=L-4), T=4, $T_0$=1, S - band {λ=4, μ=L-2} and T=2, $T_0$=1; with the constraint N=2L imposed on all the bands [5].

In Figure 2, the calculated and experimental energies are compared. The values of the parameters for all bands are given below are:

a = 0.197,
$\alpha_1$ = - 0.041690312,
$\alpha_3$ = - 0.0544336492,
$\beta_3$ = $\beta_{30}$/(1+nx),
$\beta_{30}$ = 0.23,
x = 0.0013871.

Ground band moment of inertia is determined by n=0, s-band has n=9 and for the octupole band n = 3. γ-band is constructed with two crossing bands, one of them "belongs" to the ground band n=1, another one to n =6.

It is necessary to point out that using of the same values of parameters for all the bands under consideration produces a good agreement between theory and experiment with the average deviation of the calculated energies from the experimental values for all bands together being only 28.9 keV. This agreement arises from taking into account the vibrational degrees of freedom in the determination of the moment of inertia, as was done in [5].

# ODD-EVEN STAGGERING

In our work we use the odd-even staggering function between the band states that is defined as $\Delta^5 E(L)=6\Delta E(L)-4\Delta E(L-1)-4\Delta E(L+1)+\Delta E(L+2)+\Delta E(L-2)$,
where $\Delta E(L)=E(L)-E(L-1)$ is the difference between the energies of the levels with odd and even momenta. This function checks the deviations between calculated and experimental energies and also the positions of the bands with respect to each other. In Figure 3 we present the experimental and calculated values of the function $\Delta^5 E(L)$. One can see a good agreement of our calculations with experiment.

This work was supported in part by the Bulgarian Science Foundation under contract Ф-905 and by the RFBR.

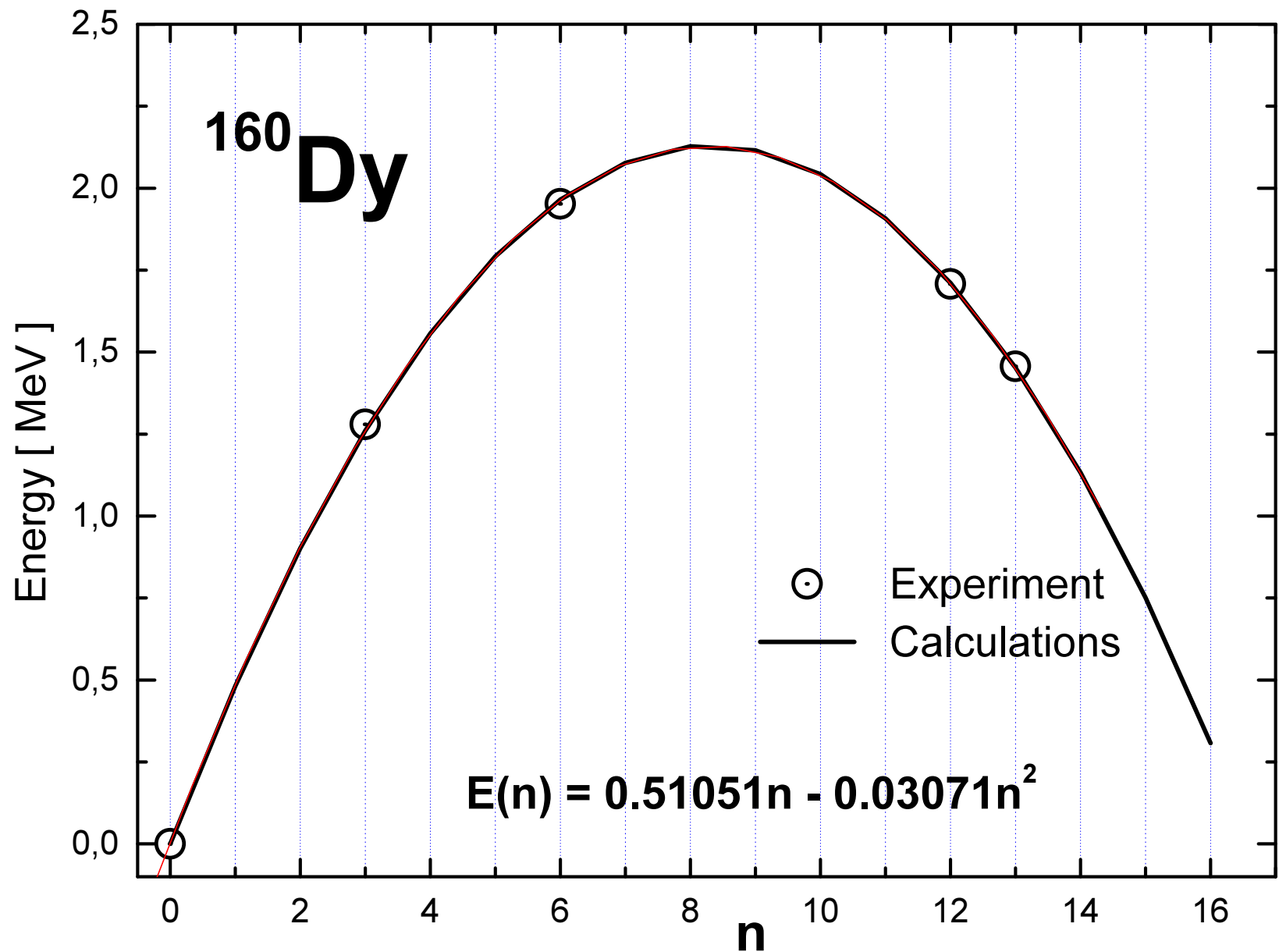

Figure 1a. The distribution of $0^+$ excited states energies.

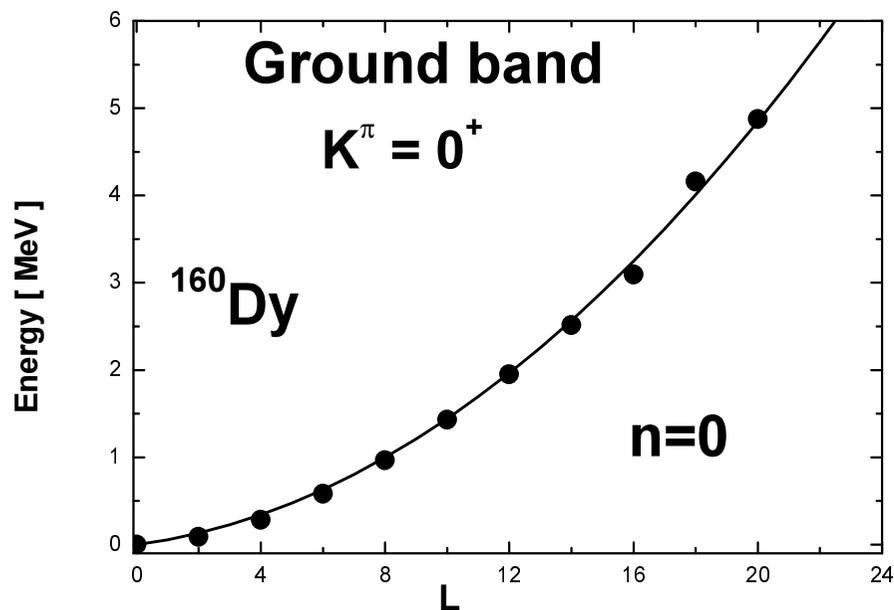
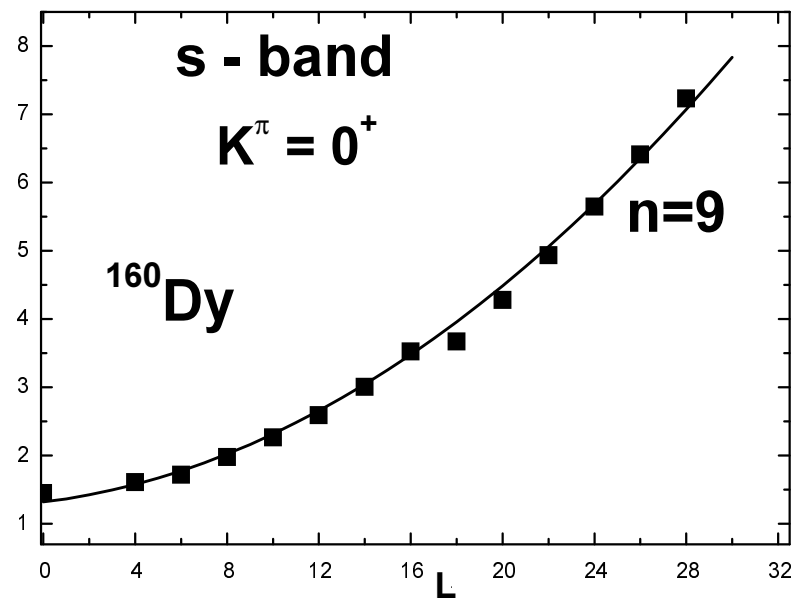
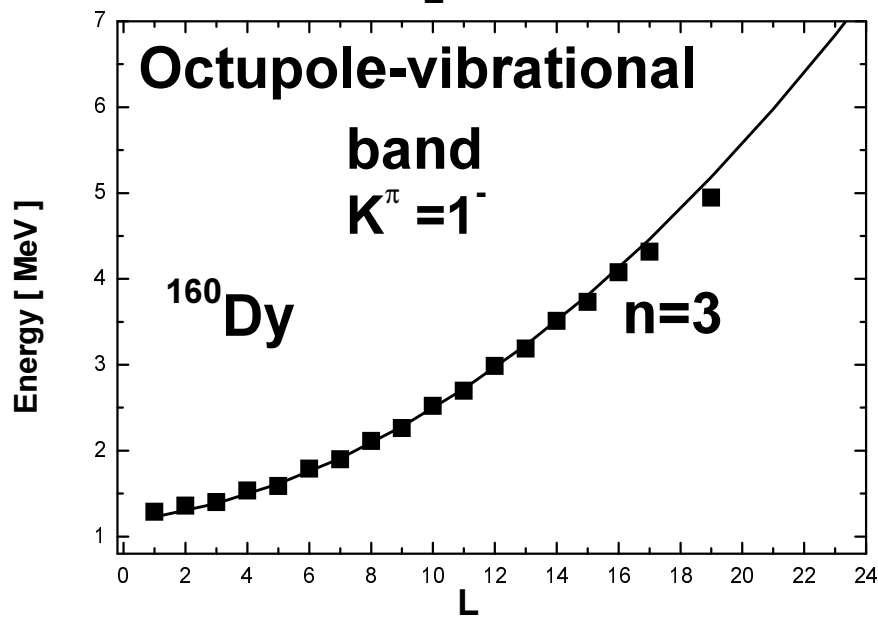
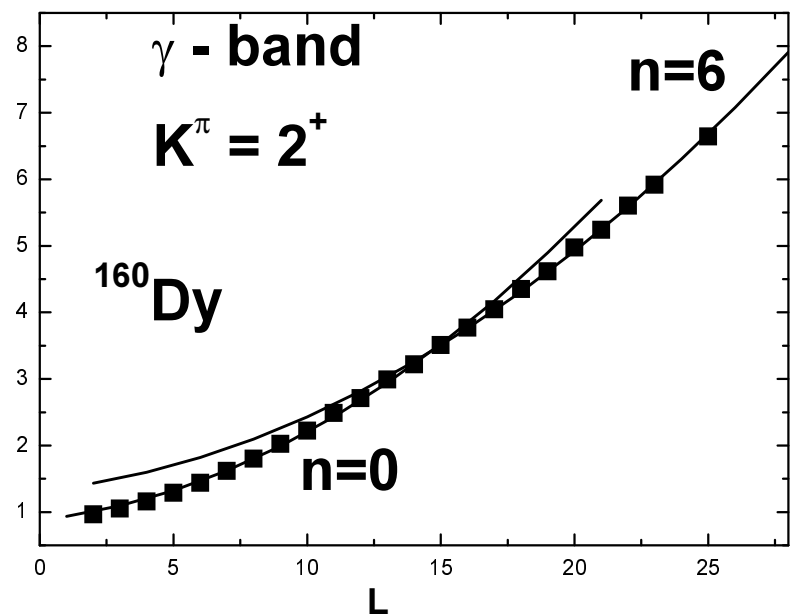

Figure 2. Comparison of IVBM calculations with experiment for different bands in $^{160}$Dy

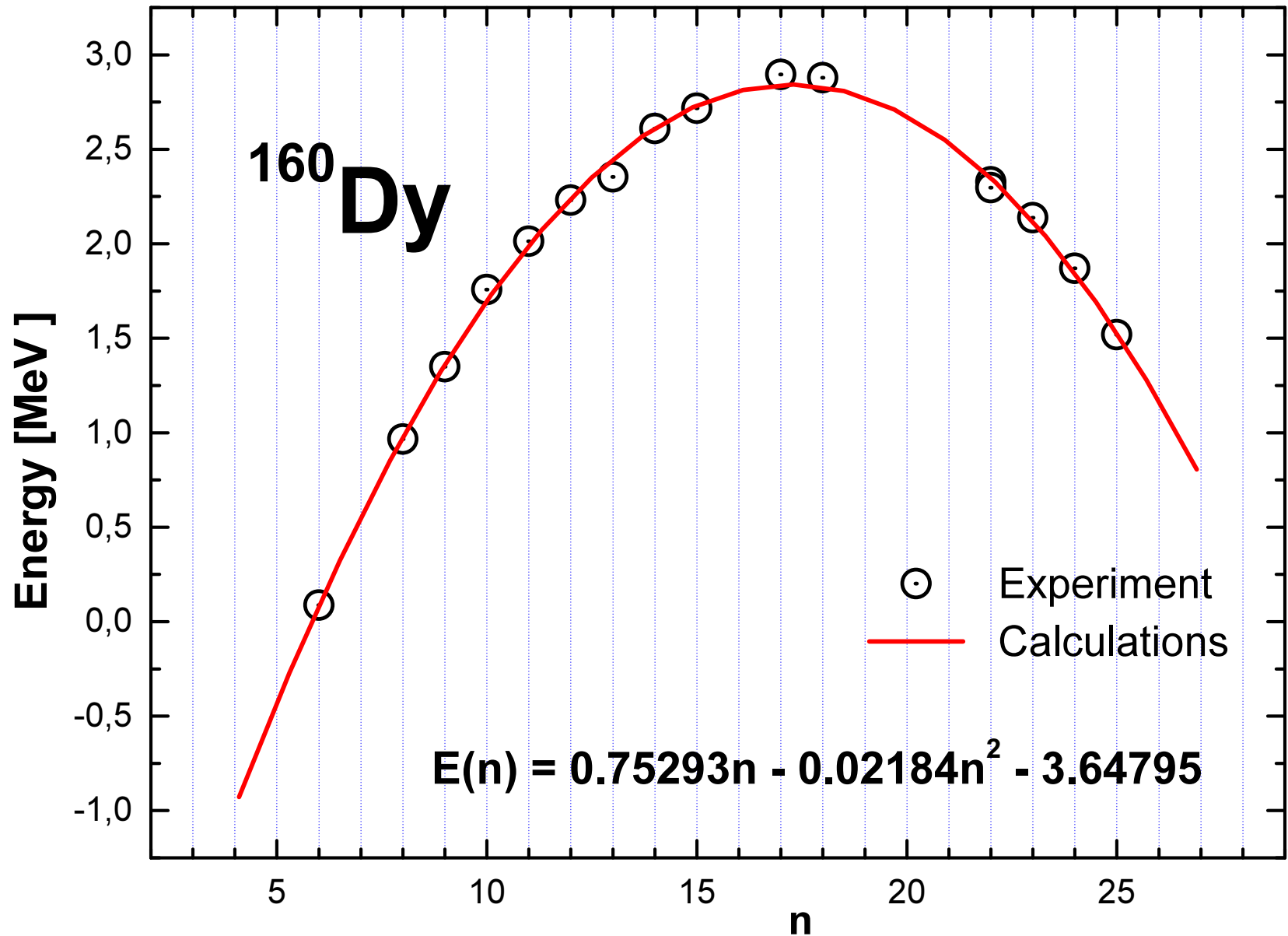

Figure 1b. The distribution of the 2⁺ excited states energies.

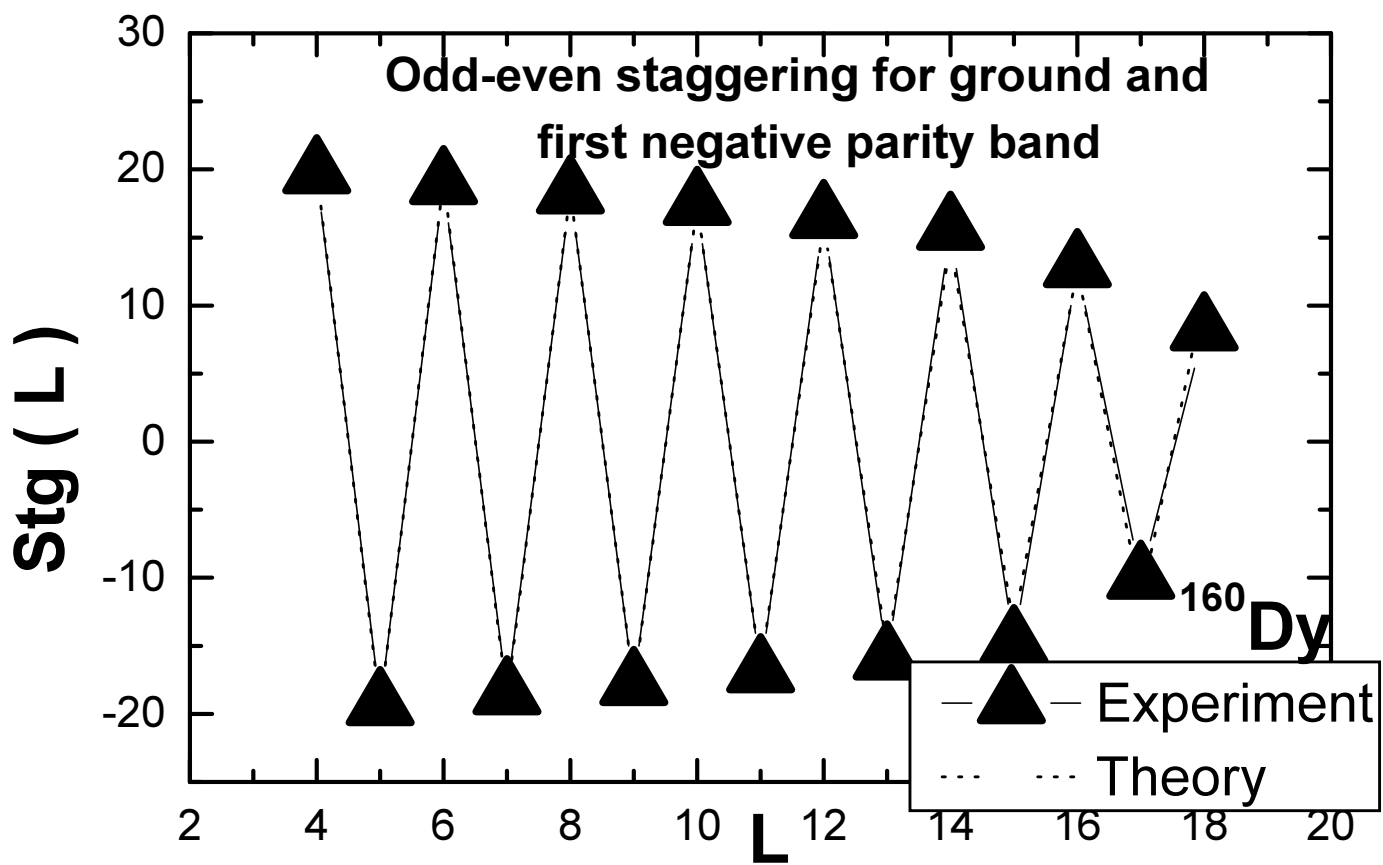

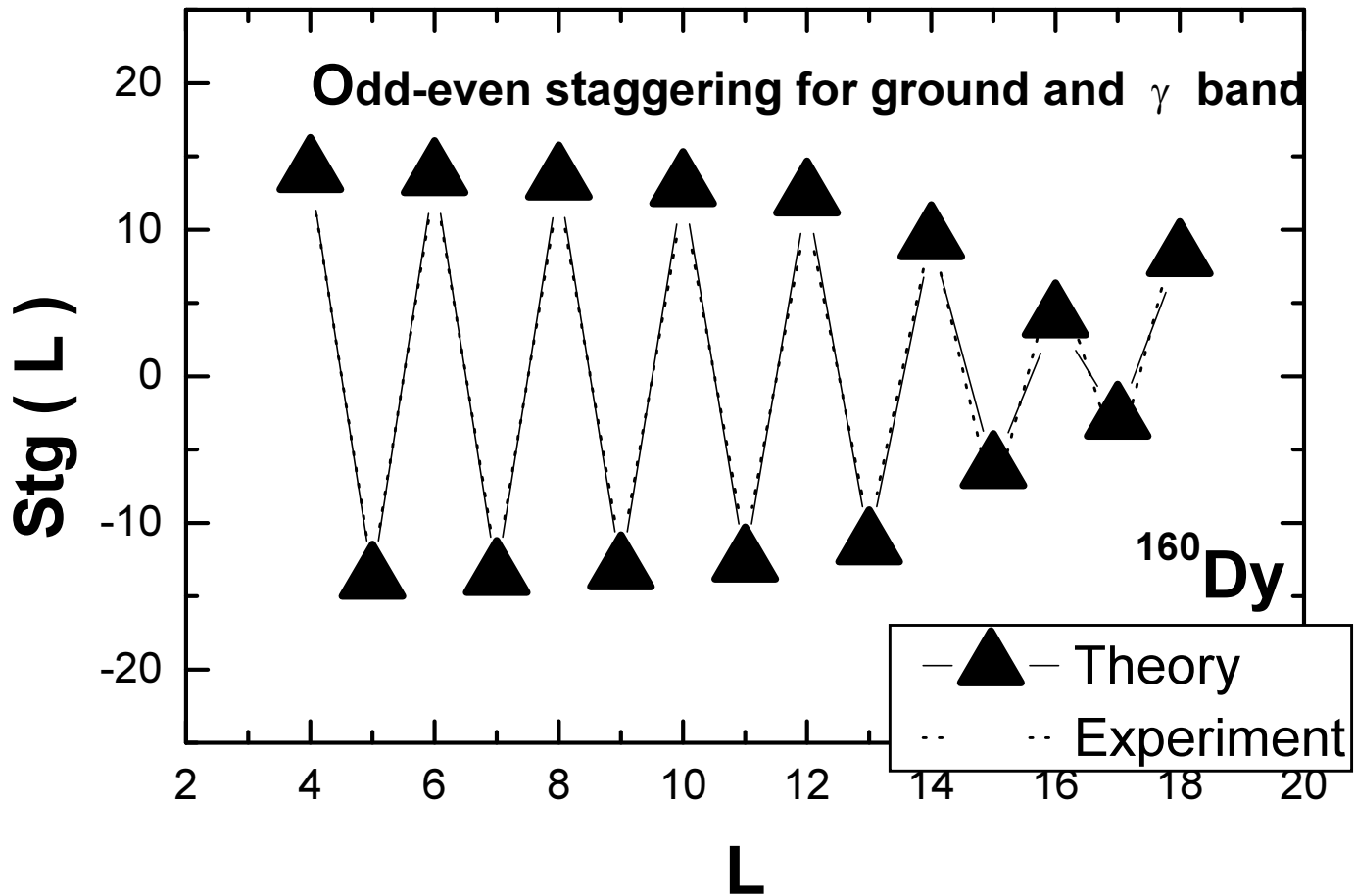

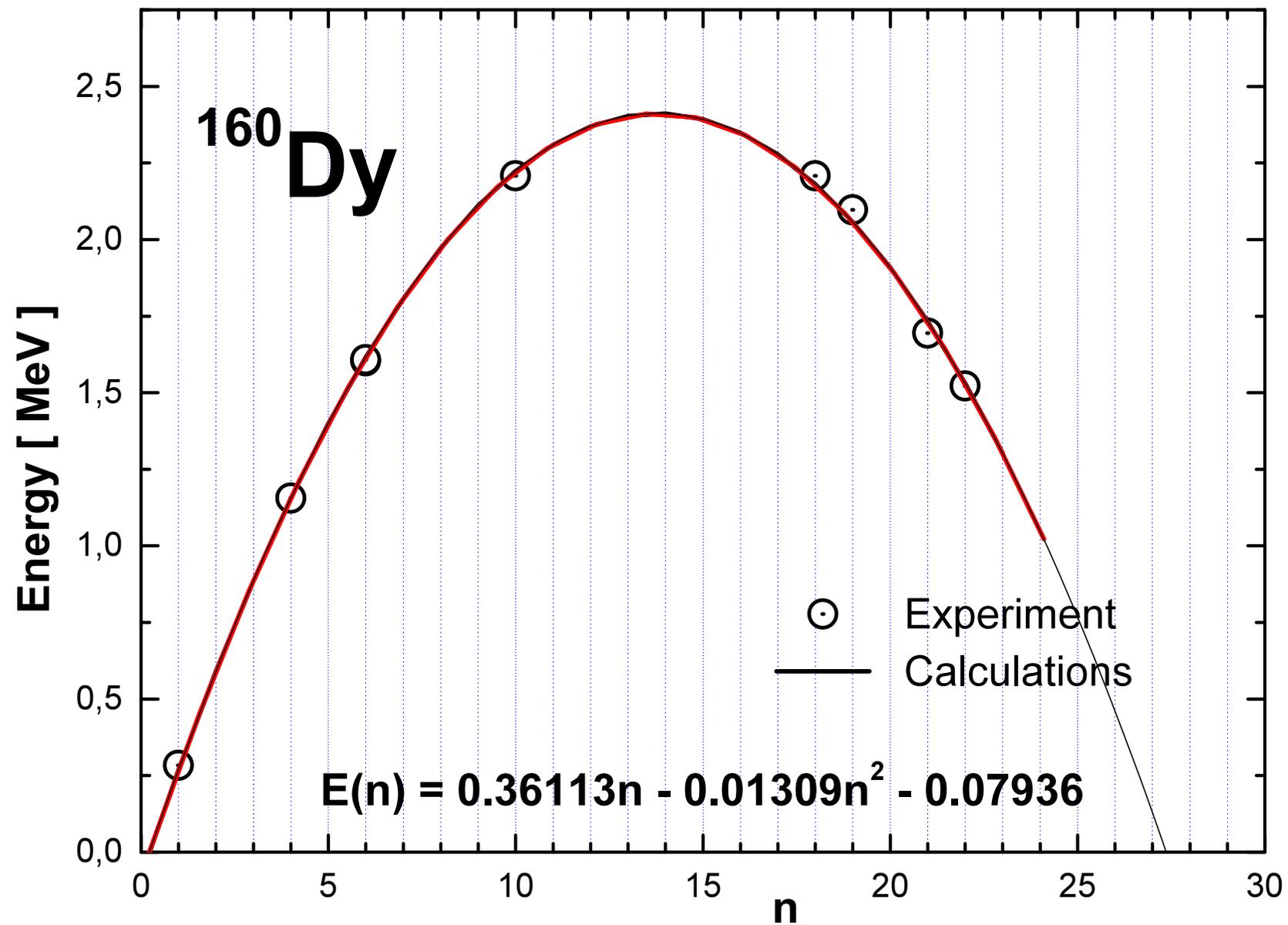

Figure 1c. The distribution of $4^+$ excited states energies.

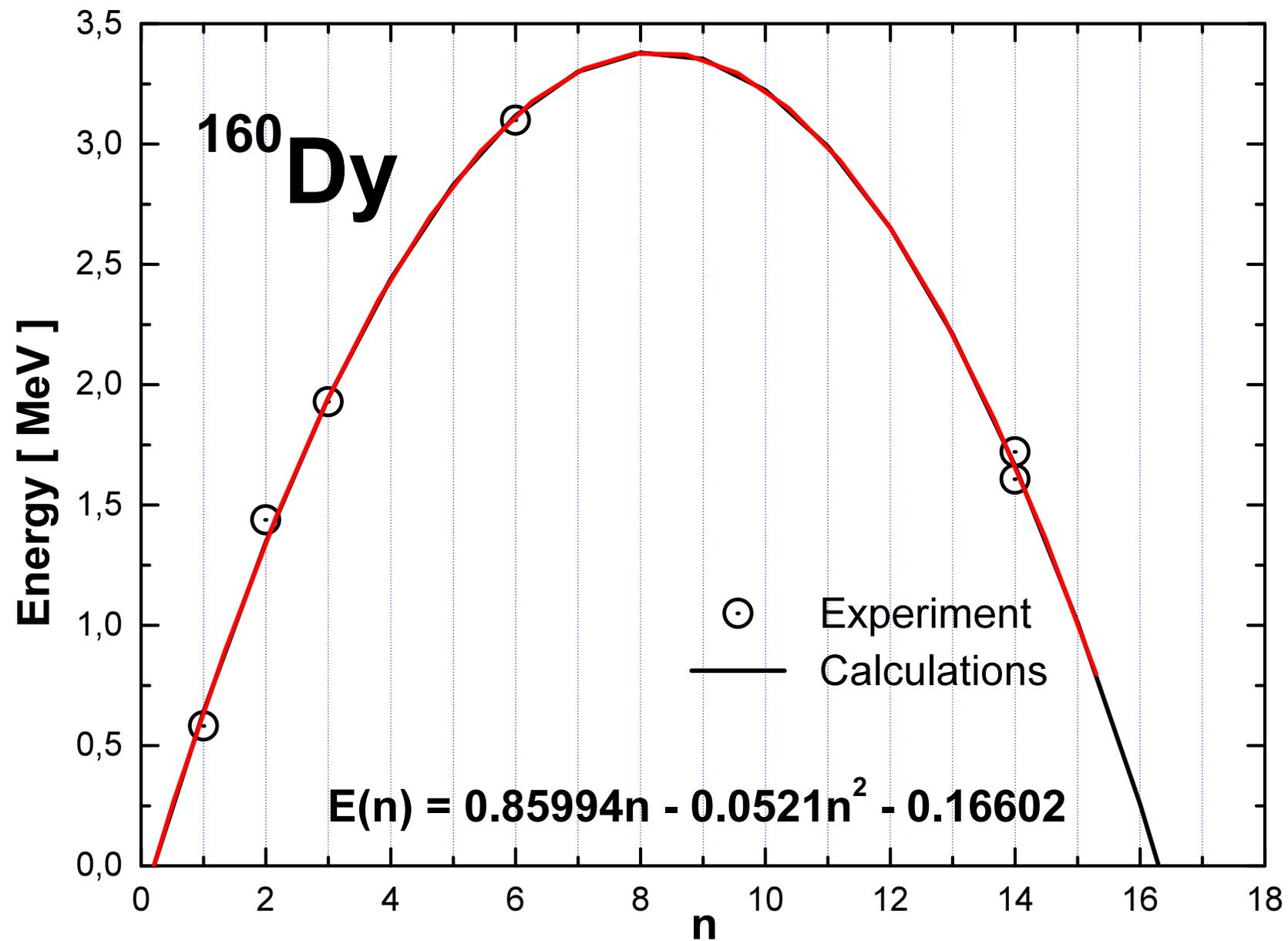

Figure 1d. The distribution of $6^+$ excited states energies.

# Decay scheme $^{160}$Er → $^{160m,g}$Ho → $^{160}$Dy

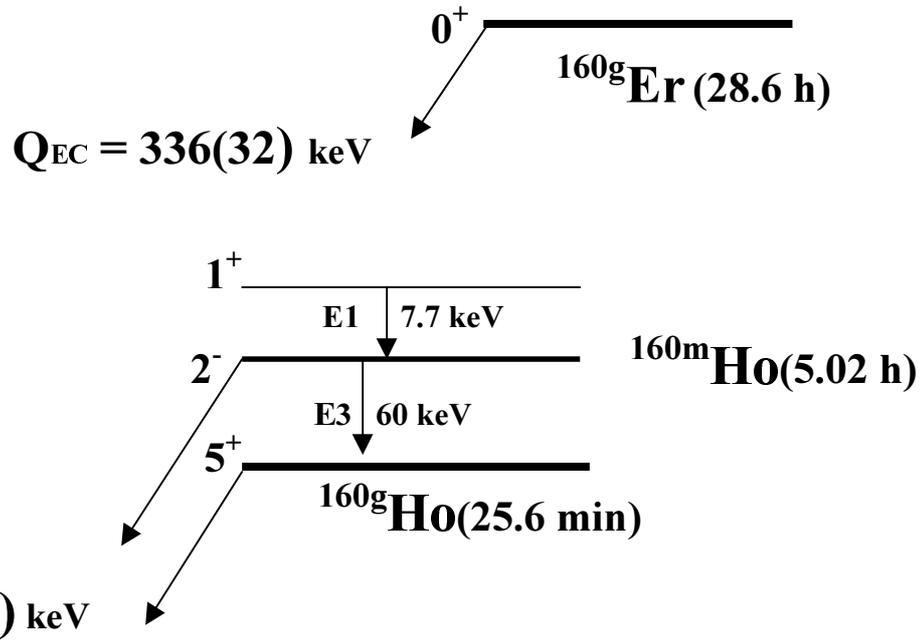

**EXCITED STATES NUMBER:** 164
**γ transitions number:** 878

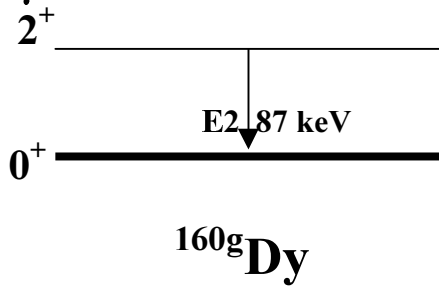